\def \< {\left\langle}
\def \> {\right\rangle}
\def \[ {\left[ }
\def \] {\right] }
\def \( {\left(}
\def \) {\right)}

\par\noindent{\bf Inertial mass and the quantum vacuum fields}
\bigskip\noindent
{\bf Bernard Haisch}
\medskip\noindent
Solar and Astrophysics Laboratory, Dept. H1-12, Bldg. 252
\par\noindent
Lockheed Martin, 3251 Hanover Street, Palo Alto, California 94304
\par\noindent
and
\par\noindent
California Institute for Physics and Astrophysics, 366 Cambridge Ave., Palo Alto, CA
94306
\par\noindent
{\it haisch@calphysics.org}
\bigskip\noindent
{\bf Alfonso Rueda}
\par\noindent
Department of Electrical Engineering, ECS Building
\par\noindent
California State University, 1250 Bellflower Blvd.,
Long Beach, California 90840
\par\noindent
{\it arueda@csulb.edu}
\bigskip\noindent
{\bf York Dobyns}
\par\noindent
C-131 Engineering Quad, Princeton Univ., Princeton, NJ 08544-5263
\par\noindent
{\it ydobyns@princeton.edu}
\bigskip

\centerline{({\it  Annalen der Physik} --- Received: 2 June 2000, accepted 12
September 2000)}

\bigskip\noindent
{\bf Abstract} --- Even when the Higgs particle is finally detected, it will continue to be a
legitimate question to ask whether the inertia of matter as a reaction force opposing
acceleration is an intrinsic or extrinsic property of matter. General
relativity specifies which geodesic path a free particle will follow, but geometrodynamics has
no mechanism for generating a reaction force for deviation from geodesic motion.
We discuss a different approach involving the electromagnetic zero-point field (ZPF) of the
quantum vacuum. It has been found that certain asymmetries arise in the ZPF as
perceived from an accelerating reference frame. In such a frame the Poynting vector and momentum
flux of the ZPF become non-zero. Scattering of this quantum radiation by the quarks and
electrons in matter can result in an acceleration-dependent reaction force. Both the ordinary
and the relativistic forms of Newton's second law, the equation of motion, can be derived from
the electrodynamics of such ZPF-particle interactions. Conjectural arguments are given why this
interaction should take place in a resonance at the Compton frequency, and how this could
simultaneously provide a physical basis for the de Broglie wavelength of a moving particle. This
affords a suggestive perspective on a deep connection between electrodynamics, the origin of
inertia and the quantum wave nature of matter.
\bigskip

\bigskip

\bigskip\noindent
{\bf 1 \ \ Introduction}
\bigskip
Although the Standard Model is customarily described as one involving fundamental particles
(leptons and quarks) and their interactions via bosons, at a deeper level it is believed that
fundamental particles are really excitations of a field. That is thought to be why all
fundamental particles of a given type, e.g. all electrons, are precisely identical. The study of
particles from this perspective, the discipline known as quantum field theory, is both
conceptually rich and quantitatively successful: witness the agreement between theory and
experiment of the magnetic moment of the electron to thirteen significant figures. Although the
technique that has been used so far to develop the hypothesis connecting inertia and the quantum
vacuum is a semi-classical one (stochastic electrodynamics), the objective is congruent with
that of quantum field theory: we are seeking an origin of inertia based on the properties of a
quantum field, although thus far we have considered only the zero-point photon field. In his
book ``Concepts of Mass in Contemporary Physics and Philosophy'' Jammer correctly states about
this approach: [1]

\medskip
{\narrower \noindent
``However, debatable as their theory still is, it is from the philosophical point of view a
thought-provoking attempt to renounce the traditional priority of the notion of mass in the
hierarchy of our conceptions of physical reality and to dispense with the concept of mass in
favor of the concept of field. In this respect their theory does to the Newtonian concept of
mass what modern physics has done to the notion of absolute space: As Einstein once wrote, `the
victory over the concept of absolute space or over that of the inertial system became possible
only because the concept of the material object was gradually replaced as the fundamental
concept of physics by that of the field.'~''

}\medskip

There was still another shift in foundation that came along with relativity, one that can be
described as the introduction of an {\it epistemology of observables}. In his 1905 paper ``On the
Electrodynamics of Moving Bodies'' Einstein eliminated the notions of a mechanical ether and of
an absolute frame of rest [2]. A consequence of his  resulting principle of relativity was the
abandonment of the concepts of absolute space and of absolute time. We see that the same dictate
of empiricist philosophy which would later (c. 1925) characterize the foundations of so-called
modern quantum theory, with its emphasis on observables, was already present in the foundations
of special relativity, though in a less overt, more concealed form. The new mechanics of
relativity which replaced that of Newton brought with it a subtle epistemological change in
foundation: relativity is founded ultimately on physically measureable quantities determined by
light propagation rather than on abstract concepts such as absolute space and absolute time. It
is the observation of light signals that defines the lengths of rulers and durations of time
intervals. Twenty years later a similar emphasis on observable or measurable quantitites became
the basis of the standard interpretation of quantum mechanics. We propose that such an 
epistemology of observables may also be appropriate for the interpretation of the concept of
mass.
\footnote{$^a$}
{A lucid discussion concerning the epistemology of observables is found in Phillip Frank's
``Einstein, Mach and Logical Positivism'' [3]. The influence on the early work of Einstein (up
to approximately 1920) by Mach and his Logical Positivistic viewpoint is widely known. The
emphasis on observables as the essence of scientific verification was widely promoted by the
thinkers of the Vienna Circle and by Auguste Compte.}

The existence of matter is self-evident and fundamental: we are made of
matter. Mass however --- like absolute space and time --- can be viewed as an abstraction. Though
it is usually regarded as an innate property of matter, mass is not in fact directly observable.
The mass we habitually attribute to matter manifests in two ways: through a force and as energy.
In classical mechanics, one applies a force,
${\bf f}$, to an object and measures its resultant acceleration, {\bf a}. The force and the
acceleration are the observables. We relate these two observables by assuming the
existence of an innate property of matter known as inertial mass and thus we write ${\bf
f} = m {\bf a}$. Viewed this way, the existence of an innate inertial mass, $m$, is an inference
and an abstraction. Using the methodology of stochastic electrodynamics [4] it has been shown
that it {\it may} be possible to view Newton's equation of motion,
${\bf f} = m {\bf a}$, as well as
its relativistic generalization,
${\cal F}=d{\cal P}/d \tau$, as a consequence of the actions of the  
electromagnetic zero-point field (ZPF) --- or more generally of the quantum vacuum fields --- on
matter [5] [6]. Of course the
situation is more complex than we have captured in our limited SED
approach, in that the quantum vacuum contains zero-point oscillations of
all gauge fields, not only electromagnetism.

With this caveat in mind that we have so far considered only the electromagnetic
quantum vacuum-matter interaction, the resistance to acceleration traditionally attributed to the
existence of inertial mass in matter appears to be logically and quantitatively attributable
instead to a resistance on accelerated matter due to the zero-point vacuum fields. In other
words, inertia would appear to be a kind of reaction force that springs into existence out of
the quantum vacuum whenever acceleration of an object takes place, for reasons given below. The
$m$ in ${\bf f} = m {\bf a}$ thus would become a coupling parameter that quantifies a more
fundamental relationship between the elementary charged particles (quarks and electrons) in
matter and the surrounding vacuum. This is not inconsistent with the ordinary concepts of
momentum and kinetic energy which are calculated using the same $m$. After all, momentum and
kinetic energy of a moving object can take on any value depending on the relative motion of the
observer, and so cannot be regarded as in any sense absolute. It is only changes, not definite
values, in momentum or kinetic energy that manifest as real measureable effects when a collision
or a mechanical interaction takes place. We would argue that momentum and energy are real
characteristics of the quantum vacua, but that for material objects momentum, energy and mass
should be viewed as calculational devices useful for predicting what will be observed when
collisions or other interactions take place. The apparent momentum, energy and mass of material
objects stem from interactions with the quantum vacua.

Our attempts to link inertia to the actions of the quantum vacua have been limited to the
electromagnetic zero-point field. We have not considered the vacua of the weak or
strong interactions.
(A recent proposal by Vigier [7] that there is also a contribution to inertia from the Dirac
vacuum goes along similar lines.) In the electromagnetic case, the inertia connection comes about
through the Poynting vector of the ZPF: in an accelerating reference frame the Poynting vector
becomes non-zero and proves to be proportional to acceleration.
\footnote{$^b$}{In this respect, the fact that here we deal with a vector field that has a
Poynting vector and not with a scalar field may be critical. For simple scalar fields such a
resistance opposing accleration is not present. This has been reviewed and studied by, e.g., 
Jaekel and Raynaud [8]. Here however [6] we are dealing with a vector field with a well
defined Poynting vector and associated momentum density}
A non-zero Poynting vector implies a
non-zero radiative momentum flux transiting any accelerating object. If one assumes that the
quarks and electrons in such an object scatter this radiation, the semi-classical
techniques of stochastic electrodynamics show that there will result a reaction force on that
accelerating object having the form
${\bf f}_r=-\mu {\bf a}$, where the $\mu$ parameter quantifies the strength of
the scattering process. In order to maintain the state of acceleration, a motive force ${\bf f}$
must continuously be applied to balance this reaction force ${\bf f}_r$. Applying Newton's third
law to the region of contact between the agent and the object, ${\bf f}=-{\bf f}_r$, we thus
immediately arrive at ${\bf f}=\mu {\bf a}$, which is identical to Newton's equation of
motion. However now a parameter originating in the zero-point field scattering,
$\mu$, accomplishes the very thing that inertial mass, $m$, is assumed to do: resist
acceleration. One can conceptually replace inertial mass, $m$, by a
ZPF-based parameter representing a scattering process,
$\mu$.
It is conceptually quite important to note that the inertial parameter
$\mu$, unlike $m$, is not an intrinsic property of the body but combines
intrinsic properties (such as $\Gamma_z$ or $\eta(\omega)$, see below) with extrinsic
parameters of the ZPF.
We discuss this relationship in $\S$ 4.

This is not merely a trivial substitution of nomenclature: {\it Taking this approach one may be
able to eliminate a postulate of physics} provided that it is eventually possible to extend this
to the strong- and weak-interaction vacuum fields. Newton's second law, ${\bf f}=m{\bf a}$, may
then cease to be fundamental as it might be derived from the vacuum fields plus the third law.
Newton's third law of action and reaction would be axiomatic; Newton's second law would not.
For  practical purposes one could retain the concept of inertial mass,
$m$, while realizing that it is an
emergent property of matter/field interactions. One might regard mass in the
same category as a classical thermodynamic parameter, such as heat capacity, for
example. The measurable heat capacity of a given substance is a useful concept, but we know
that it really represents an ensemble of atomic processes at a more fundamental level. So
it appears to be with inertial mass as well.

In conventional QCD the proton and neutron masses are explained as being primarily the energies
associated with quark motions and gluon fields, the masses of the $u$ and $d$ quarks amounting
to very little (app. 20 MeV in comparison to a nucleon mass of about 1 GeV). That sort of
reasoning --- calculating binding and kinetic energies, etc. --- is usually considered sufficient
explanation of nucleon masses, but the quantum vacuum-inertia hypothesis addresses the
possibility that there is a deeper level to the nature of mass {\it by asking where inertia
itself comes from}. In other words, even if QCD calculations yield a correct
mass-equivalent energy for a nucleon, can one still ask the question why that energy possesses
the property of resisting acceleration? We are proposing that there may be such a physical basis
underlying the reaction force that characterizes inertia. If this is true, that would certainly
be a deeper explanation than simply saying that there is so much energy (mass) in the quark
motions and gluon fields and by definition that such energy (mass) simply resists acceleration.
Where does the specific reaction force that opposes acceleration come from? Why does mass
or its energy equivalent resist acceleration? One possibility is that  this will never be solved
and forever remain a mystery. Another possibility is that this {\it can} be explained and that
the present approach  offers a possible new insight.

Inertial mass is only one of several manifestations of the concept of mass. If a
ZPF-scattering process can account, at least in part, for inertial mass
is there an analogous basis for the $E=mc^2$ relation? This equation is often seen
as a statement
that one kind of thing (energy) can be transformed into a totally different kind of thing (mass)
and vice versa, but this is probably a misleading view.  Most of the derivations of
this relationship show that it is really a statement that {\it energy has
inertia\/}; that Newton's second law could be rewritten ${\bf F} =
(E/c^2){\bf a}$.[1] Indeed, it seems likely that this relationship follows
automatically regardless of the detailed theory (if any) of inertia, so
long as that theory maintains the conservation of energy and momentum,
because those conservation laws (together with the axioms of special
relativity) are sufficient to determine how the inertia of a body changes
when it emits or absorbs energy.

Following an epistemology of
observables, we propose that this is indeed the case, and that just as inertial mass may be
regarded as an abstraction postulated to account for the observation of an
acceleration-dependent force, rest mass may be an abstraction accounting for some kind of
internal, ZPF-based energy associated with the fundamental particles constituting matter. In
a preliminary attempt to develop the Sakharov [9] conjecture of a vacuum-fluctuation model for
gravity, Hestenes[10] proposed that the
$E=mc^2$ relationship reflected the internal energy associated with {\it
zitterbewegung} of fundamental particles (see also Puthoff [11] for a similar suggestion).
The {\it zitterbewegung}, so named by Schr\"odinger [12], can be understood as the
ultrarelativistic oscillatory motions associated with the center of charge operator in the
electron with respect to the center of mass operator. It can be interpreted as a motion of the
center of charge around the averaged center of mass point. It is attributed in stochastic
electrodynamics to the fluctuations induced by the ZPF. In the Dirac theory of the electron the
eigenvalues of the {\it zitterbewegung} velocity are
$\pm c$ (see [13]), and the amplitude of these oscillations are on the order of the Compton
wavelength.  In the view proposed by Schr\"odinger, Huang, Hestenes and others, the rest mass of
a particle is actually the field energy associated with point charge particle oscillations driven
by the ZPF. If that is the case, there is no problematic conversion of mass into energy or
enigmatic creation of mass from energy, but rather simply a concentration or liberation of
ZPF-associated energy. Here too mass may become a useful but no longer fundamental concept.

{\it This approach may allow yet another reduction in physical postulates.}
Just as the laws of electrodynamics applied to the ZPF appear to explain and support a former
postulate of physics (${\bf f}=m {\bf a}$) via a new interpretation of inertial mass, a
postulate of quantum mechanics appears to be derivable via an interpretation of rest mass as
the energy of ZPF-driven {\it zitterbewegung}: The de Broglie relation for the wavelength of a
moving particle, $\lambda_B=h/p$, may be derived from Doppler shifts of the Compton-frequency
oscillations associated with {\it zitterbewegung} that occur when a particle is placed in
motion. This is discussed in
$\S$ 5.

There is one final mass concept: gravitational mass. Einstein's principle of equivalence dictates
that inertial and gravitational mass must be the same. Therefore if inertial mass is a
placeholder for vacuum field forces that arise in accelerating reference frames, then
there must be an analogous connection between gravitation and vacuum fields.
The attempt of Puthoff over a decade ago to develop the Sakharov conjecture along the lines of a
stochastic electrodynamics approach was stimulating, but has not yet been successful in
accounting for Newtonian gravitation.[14] We limit our discussion on gravitation to some
comments on this and on the associated problem of the cosmological constant in
$\S$ 6.

To summarize the view that emerges from these considerations, all energy and momentum that we
normally associate with matter may actually reflect some part of the energy and momentum of the
underlying vacuum. The classical kinetic energy, $T=mv^2/2$, or momentum, ${\bf p}=m{\bf v}$,
that we ascribe to an object depend entirely on the relative motion of the object and the
observer. Both $T$ and {\bf p} are necessarily calculated quantities; a real observation only
arises when object and observer are made to closely interact, e.g. when brought together into
the same frame, which is to say when a collision occurs. But to achieve that requires a change in
velocity, and it is precisely upon deceleration that the vacuum generates a reaction force that
is called the inertial reaction force which Newton took to be an irreducible property of the
so-called inertial mass,
$m$. Again, we may retain the concept of inertial mass as a convenient bookkeeping tool for
kinetic energy, momentum and other calculations, but the actual observable measurement of forces
can perhaps, we are suggesting, be traced back to the vacuum reaction force on the most
elementary components of matter (e.g., in the electromagnetic case, quarks and electrons) that
accompanies acceleration.

\bigskip
\bigskip\noindent
{\bf 2 \ \ Historical remarks on the zero-point field of Stochastic Electrodynamics}
\bigskip

Any physical field must have an associated energy density; therefore
the average field intensity over some small volume is associated with 
a given energy. The Heisenberg uncertainty relation (in the $\Delta E
\Delta t$ form) requires that this energy be uncertain in inverse
proportion to the length of time over which it obtains. This uncertainty
requires fluctuations in the field intensity, from one such small volume
to another, and from one increment of time to the next; fluctuations which
must entail fluctuations in the fields themselves. These fluctuations become
more intense as the spatial and temporal resolution increases. 

Such quantum arguments apply to the electromagnetic field. The quantization of the
field in terms of quantum-mechanical operators may be found in various standard textbooks, such
as that of Loudon [15]: ``The electromagnetic field is now quantized by the association of a
quantum-mechanical harmonic oscillator with each mode {\bf k} of the radiation field."
This can easily be understood: Application of the Heisenberg uncertainty relation to a harmonic
oscillator immediately requires that its ground state have a non-zero energy of $h\nu/2$,
because a particle cannot simultaneously be exactly at the bottom of its potential well and have
exactly zero momentum. The harmonic oscillators of the EM field are formally identical to those
derived for a particle in a suitable potential well; thus there is
the same $h\nu/2$ zero-point energy expression for each mode of the
field as is the case for a mechanical oscillator.  Summing up the energy
over the modes for all frequencies, directions, and polarization states,
one arrives at a zero-point energy density for the electromagnetic
fluctuations, and this is the origin of the electromagnetic ZPF.
An energy of $h\nu/2$ per mode of the field characterizes both the fluctuations of the quantized
radiation field in quantum field theory and the amplitude of random electromagnetic plane waves
in stochastic electrodynamics.

The clearest introduction to the classical electromagnetic ZPF concept of Stochastic
Electrodynamics (SED) was the review paper of Boyer in 1975 [16] that discussed the foundational
aspects of SED theory. In the Lorentz-Maxwell classical electrodynamics or Lorentz theory of the
electron, one automatically assigns a zero value everywhere for the homogeneous solutions
for the potential equations. In other words, it is taken for granted that the classical electron
is not immersed in an incoming free background field: all electromagnetic radiation at any point
in the Universe is due solely to discrete sources or to the remnant radiation from the Big Bang.
Boyer argued that this is not the only possible assumption: it is also legitimate to assume a
completely random but on average homogeneous and isotropic electromagnetic radiation field
provided that it has an energy density spectrum that is Lorentz invariant. If this is so,
identical experiments will yield exactly the same results when performed in totally different
inertial frames because the spectrum looks the same from all inertial frames. {\it Since it
consists of ordinary electrodynamics, the ZPF of SED is by definition consistent with special
relativity.} It was shown by Marshall [17] and later independently by Boyer that the only
spectrum of a random field with these characteristics is a
$\nu^3$ distributed spectral energy density. This is exactly the form of the spectrum studied by
Planck in 1911 [18] and is the spectrum of the ZPF that emerges from QED. This ZPF is not related
to the 2.7 K cosmic microwave remnant radiation of the Big Bang.

SED is thus precisely the Lorentz classical electrodynamics with the sole addendum of a
uniform, isotropic, totally random electromagnetic radiation field (the ZPF) having a $\nu^3$
spectral energy density whose value is scaled by Planck's constant, $h$.
In this view, $h$ is not a unit of quantization nor quantum of action, but rather a scaling
parameter for the energy density of the ZPF.

One rationale of SED has been to explore
a possible classical foundation for quantum fluctuations, which in this view, may be interpreted
as the result of random electromagnetic perturbations; for that reason $h$ as a measure (or
degree) of quantum uncertainty translates into a measure or scale of ZPF energy density in SED
since electromagnetic fluctuations are assumed to generate uncertainty as embodied in the
Heisenberg relation in the conventional quantum view.

Another rationale of SED, and by far the most useful one in our view, is as a powerful and
intuitive calculational tool for certain kinds of problems. We do not seriously expect, as other
SED researchers appear to, that SED in any substantial way may replace or supplant quantum
theory. However it is not unreasonable to expect that SED theory may throw some light into
foundational aspects of quantum theory. One revealing outcome of SED has been that some aspects
of quantum mechanics would appear to be explicable in terms of classical
electrodynamics if one accepts as an {\it Ansatz} the existence of a real electromagnetic ZPF.
Another outcome has been the use of its techniques for the predictions or explanations of some
effects that so far have remained unexplained.

Planck [18] derived a closed mathematical expression that fit the measurement of the
spectral distribution of thermal radiation by hypothesizing a quantization of the radiation
emission process. This yielded the well-known blackbody function,

$$\rho(\nu,T) = {8\pi\nu^2 \over c^3}
\left( {h\nu \over e^{h\nu/kT}-1} \right) ,
\eqno(1)$$

\smallskip\noindent written here as an energy density and factored so as to show
the two components: a density of modes (i.e. number of degrees of freedom per unit
volume) times the thermal energy per mode in the frequency interval $d\nu$. As discussed in
detail in Kuhn [19], Planck himself remained skeptical of the physical significance and
importance of his theoretical discovery of an apparently new constant of nature, $h$, for
over a decade.

In 1913 Einstein and Stern [20] studied the interaction of matter with
radiation using classical physics and a model  of simple dipole oscillators to represent charged
particles.  They found that if, for some reason, such a dipole oscillator had a zero-point
energy, i.e. an irreducible energy even at $T=0$, of $h\nu$, the Planck formula for the 
radiation spectrum would result {\it without the need to postulate quantization as an a priori
assumption.}

The existence of such a ZPF had already been envisaged  by Planck around 1910 when he
formulated his so-called second theory:   namely an attempt to derive the blackbody spectral
formula with a  weaker quantization assumption.  Nernst [21] proposed that  the Universe
might actually contain enormous amounts of such ZPF  radiation and became the main proponent
of this concept.  Both Planck and Nernst used the correct $h\nu/2$ form for the average 
energy of the zero-point electromagnetic fluctuations instead of the $h\nu$  value assumed by
Einstein and Stern; the $h\nu$ assumption is correct for  the sum of interacting harmonic
oscillator plus the energy of the  electromagnetic field mode. The electromagnetic blackbody
spectrum including ZPF would then be:

$$\rho(\nu,T) = {8\pi\nu^2 \over c^3}
\left( {h\nu \over e^{h\nu/kT}-1} + {h\nu \over 2} \right) .
\eqno(2)$$

\smallskip\noindent
This appears to result in a $\nu^3$ ultraviolet catastrophe in the
second term. In the context of SED, however, that divergence is not fatal.
This component now refers not to measurable excess radiation from a heated object, but
rather to a uniform, isotropic background radiation field that cannot be directly measured
because of its homogeneity and isotropy. This approach of Einstein and Stern to understanding the
blackbody spectrum was not  developed further thereafter, and was essentially forgotten for the 
next fifty years until its rediscovery by Marshall [17]. In recent times, several modern
derivations of the blackbody function using classical physics with a real ZPF but without
quantization (i.e. SED) have been presented mainly by Boyer (see Boyer [22] and references
therein; also de la Pe\~na and Cetto [4] for a thorough review and references to other authors).
In other words, if one grants the existence of a real ZPF, the correct blackbody formula for the
thermal emission of matter seems to naturally follow from classical physics without quantization.

Another curiousity of the SED approach is that it could have provided a different method of
attack to the problem of the stability of the ground-state of hydrogen.
Rutherford's discovery of the atomic nucleus in 1911 together with Thomson's previous
discovery of the electron in 1897 led to the analogy between atomic structure and planetary
orbits about the Sun. In this naive analogy however, electrons, being charged, would radiate
away their orbital energy and quickly collapse into the nucleus. Bohr [23] resolved the
problem of radiative collapse of the hydrogen atom.  He recognized that Planck's constant,
$h$, could be combined with Rydberg's empirical relationship among the spectral lines of
hydrogen to solve the problem of atomic stability by boldly postulating that only discrete
transitions are allowed between states whose angular momenta are multiples of $\hbar$, where
$\hbar=h/2\pi$. The ground state of the hydrogen atom would then have angular momentum
$mva_0=\hbar$, or equivalently $m\omega_0 a_0^2=\hbar$, and would be forbidden to decay below
this ``orbit'' by Bohr's fiat. A more complex picture quickly developed from this that
substituted wave functions for orbiting point particles, and in that view the orbital angular
momentum of the ground state is actually $l=0$: the wavefunction is spherically symmetric and
has a radial probability distribution whose most probable value is $a_0$ (the expectation value
being ${3 \over 2} a_0$).

As with the classical derivation of the blackbody function made possible by the assumption of
a real ZPF, modern SED analysis of the Bohr hydrogen atom has yielded a suggestive insight.
A simple argument assuming strictly circular orbits by Boyer [15] and Puthoff [24] indicated that
while a classically circularly-orbiting electron would indeed radiate away energy, if one
takes into account the ZPF as a source of energy to be absorbed, then it is at the Bohr orbit,
$a_0$, that a condition of balance would take place in absorbed and emitted power such that
$<P^{abs}>_{circ}=<P^{rad}>_{circ}$. In other words, a classically orbiting and radiating
electron would pick up as much energy as it loses, and thus be energetically stabilized. In the
analysis a strong assumption was introduced, namely that the electron moves around the nucleus
along strictly circular orbits. This stabilization was found to be somewhat at odds with the
more realistic analysis of Claverie and coworkers [25] who studied the problem in detail. A
prediction of this much more detailed stochastic but still subrelativistic analysis was that the
atom would, unfortunately, undergo self-ionization. A study has recently been initiated by D.
Cole to look more carefully at the details of the binding potentials with the
expectation that self-ionization effects might be quenched. The physical ideas underlying
this approach were developed a decade ago [26] but were, unfortunately, intractable at the time.
Numerical simulation techniques are now available to deal with the extremely non-linear
dynamics involved.

The detailed SED analysis of Claverie and
coworkers was not restricted to global quantities and contemplated the general case of orbits not
restricted to be circular, but where the much more realistic stochastic motion was allowed to
happen. It used the more sophisticated Fokker-Planck approach (see [25] and
references therein) and it involved other dynamic quantities such as momentum and not just
average energies. But, being subrelativistic, these models assumed the electron to be a purely
pointlike particle with no structure and they therefore neglected {\it zitterbewegung} and spin,
ingredients that surely are relevant and probably essential for the stability of the
hydrogen atom. This was discussed in detail by Rueda [27]; see also Haisch, Rueda and Puthoff
[28] and de la Pe\~na and Cetto [4] for a general discussion and references.
The ultrarelativistic point-electron motions should be an essential ingredient not only in the
constitution of the particle itself but also in the stability of its states in the hydrogen atom.
This is why an SED theory at subrelativistic speeds and without possibilities to apprehend the
particle structure features is unlikely to succeed in solving problems such as that of the
stability of the hydrogen atom. The fact that $\hbar$ independently appears in the ZPF spectrum
and in the spin of the electron clearly points towards some common origin. The proper SED study
of this will require dealing not only with the difficulties of the ultrarelativistic speeds of
the electron point charge, but also with stochastic non-linear partial
differential equations with colored noise that are well beyond present-day techniques. [27]

\bigskip\noindent
{\bf 3 \ \ The zero-point field as viewed from uniformly-accelerating reference frames}
\bigskip

The ZPF spectral energy density, cf. eqn (3),

$$
\rho_{zp}(\nu)={4 \pi h \nu^3 \over c^3} \eqno(3)
$$ 
would indeed be analogous to a
spatially
uniform constant offset that cancels out when considering net energy fluxes.
However an
important discovery was made in the mid-1970s that showed that the ZPF
acquires special
characteristics when viewed from an accelerating frame.
In connection with radiation from evaporating black holes as proposed in 1974 by
Hawking [29], working independently Davies [30] and Unruh [31]
determined that a
Planck-like component of
the ZPF will arise in a uniformly-accelerated coordinate system, namely one having
a constant proper
acceleration {\bf a} with what amounts to an effective
``temperature''

$$T_a = {\hbar a \over 2 \pi c k} ,  \eqno(4)$$

\smallskip\noindent
where $a=|{\bf a}|$. This ``temperature'' does not originate in emission from particles
undergoing thermal
motions.
\footnote{$^c$}{One suspects of course that there is a deep connection
between the fact
that the ZPF spectrum that arises in this fashion due to acceleration and
the ordinary
blackbody spectrum have identical form.}
As discussed by Davies, Dray and Manogue [32]:

\medskip
{\narrower \noindent
``One of the most curious properties to be discussed in recent years is the
prediction that
an observer who accelerates in the conventional quantum vacuum of Minkowski
space will
perceive a bath of radiation, while an inertial observer of course
perceives nothing. In
the case of linear acceleration, for which there exists an extensive
literature, the
response of a model particle detector mimics the effect of its being
immersed in a bath of
thermal radiation (the so-called Unruh effect).''

}
\medskip\noindent
This ``heat bath'' is a quantum phenomenon. The ``temperature'' is
negligible for most
accelerations. Only in the extremely large gravitational fields of black
holes or in
high-energy particle collisions  can this temperature become significant. This effect
has been
studied using both QED [30][31] and the SED formalism [33]. For the classical
SED case it is found that the spectrum is quasi-Planckian
in $T_a$. Thus for the case of zero true external thermal radiation
$(T=0)$ but including this acceleration effect
$(T_a)$, eqn. (3) becomes [33]
\footnote{$^d$}
{However, further analysis by Boyer [34] showed that although the spectrum of the fields in an
accelerated frame is correctly given by eqn. (5), a dipole oscillator attached to the
uniformly-accelerated frame will have an additional radiation reaction term that exactly
compensates for the additional factor
$\left[ 1+ \left( a/2\pi c \nu \right)^2 \right]$ in eqn. (5). As a result the detector will
still detect only a Planckian spectrum insofar as the {\it scalar} detector-ZPF interaction is
concerned!}

$$\rho(\nu,T_a) = {8\pi\nu^2 \over c^3}
\left[ 1 + \left( {a \over 2 \pi c \nu} \right) ^2 \right]
\left[ {h\nu \over 2} + {h\nu \over e^{h\nu/kT_a}-1} \right] , \eqno(5)$$

\smallskip\noindent where the acceleration-dependent pseudo-Planckian
component is placed
after the $h\nu/2$ term to indicate that except for extreme accelerations
(e.g. particle
collisions at high energies) this term is negligibly small.
While these additional acceleration-dependent terms do not show any spatial
asymmetry in
the expression for the ZPF spectral energy density, certain asymmetries do
appear when the (vector)
electromagnetic field interactions with charged particles are analyzed, or
when the
momentum flux of the ZPF is calculated. The ordinary plus $a^2$ radiation
reaction terms
in eqn. (12) of HRP mirror the two leading terms in eqn. (5).

An analysis was presented by HRP and this resulted in the apparent derivation of
at least part of Newton's equation of motion, ${\bf f}=m{\bf a}$, from Maxwell's equations as
applied to the ZPF. In that analysis it appeared that the resistance to acceleration known as
inertia was in reality the electromagnetic Lorentz force stemming from interactions between a
charged particle (such as an electron or a quark) treated as a classical Planck oscillator and
the ZPF, i.e. it was found that the stochastically-averaged expression
 $<{\bf v}_{osc} \times {\bf B}^{zp}>$ was exactly proportional to and in the
opposite direction to the acceleration ${\bf a}$. The velocity
${\bf v}_{osc}$ represented the internal velocity of oscillation induced by the electric
component of the ZPF, ${\bf E}^{zp}$, on the harmonic oscillator. This internal motion was
restricted to a plane orthogonal to the external direction of motion (acceleration) of the
particle as a whole. The Lorentz force was found using a perturbation technique due to
Einstein and Hopf [35]. Owing to its linear dependence on acceleration we interpreted
this resulting force as a contribution to Newton's inertia reaction force on the particle.

The HRP analysis can be summarized as follows. The simplest possible model of a particle
(which, following Feynman's terminology, was referred to as a parton) is that of a
harmonically-oscillating point charge (``Planck oscillator''). Such a model would apply to
electrons or to the quarks constituting protons and neutrons for example. Given the peculiar
character of the strong interaction that it increases in strength with distance, to a first
approximation it is reasonable in such an exploratory attempt to treat the three quarks in a
proton or neutron as independent oscillators. This Planck oscillator is driven by the electric
component of the ZPF,  {\bf E}$^{zp}$, to motion with instantaneous velocity,
${\bf v}_{osc}$, assumed for simplicity to be in a plane perpendicular to the direction of the
externally-imposed uniform acceleration. This oscillatory-type motion is the well-known {\it
zitterbewegung} motion. The oscillator moves under constant proper acceleration,
${\bf a}$, imposed by an independent, external agent. New components of the ZPF will appear in
the frame of the accelerating particle having the spectral energy density given in eqn. (5). The
leading term of the acceleration-dependent terms is taken; the electric and magnetic fields are
transformed into a constant proper acceleration frame using well-known relations. The Lorentz
force, ${\bf f}_L$, arising from the acceleration-dependent part of the {\bf B}$^{zp}$ acting
upon the Planck oscillator is calculated: it is found to be proportional to acceleration.
Following the approach in $\S1$, our result may be expressed as, ${\bf f}_r={\bf f}_L = -m_i
{\bf a}$, i.e. the reaction force created by the ZPF due to acceleration through the quantum
vacuuum is the Lorentz force, and $m_i$ is an electromagetic parameter. To maintain the
acceleration process, a motive force, {\bf f}, must continually be applied to compensate for
${\bf f}_r$, and therefore
${\bf f}=-{\bf f}_r=m_i {\bf a}$.

The electromagnetic constant of proportionality, $m_i$, is interpreted as the physical
basis of the inertial mass of the Planck oscillator and thus at least as a contribution to the
total mass of the real particle that was modeled for simplicity as a Planck oscillator.  This
inertial mass,
$m_i$, was found to be a function of a radiation damping constant for {\it zitterbewegung},
$\Gamma_z$, of the oscillator  and of the interaction frequency with the ZPF. In the HRP
analysis it was assumed that the interaction between the ZPF and the Planck oscillator takes
place at a very high cutoff frequency, $\omega_c$, which was suggested to be the Planck
frequency, or perhaps a limiting frequency reflecting some minimum size of an elementary
particle. The expression that was found relating mass to the {\it zitterbewegung} damping
constant and a cutoff frequency was (eqn. 111 of HRP):

$$m_i = {\Gamma_z \hbar \omega_c^2 \over 2 \pi c^2} . \eqno (6)$$

\smallskip\noindent
For reasons discussed below, we now think it more likely that the ZPF-parton
interaction takes place at a resonance. We thus replace $\omega_c$, a cutoff frequency, by
$\omega_r$, a resonance frequency, and rewrite eqn. (6) in terms of cycle frequency, $\nu_r$,
rather than angular frequency,
$\omega_r$, to arrive at

$$
m_i = \Gamma_z \nu_r {h \nu_r \over c^2}. \eqno(7)
$$
For the case of the electron, if $\nu_r=\nu_C$, the Compton frequency, then $h\nu_c=512$ keV and
clearly we then want $\Gamma_z \nu_c=1$ for the mass of the electron to ``come out right.'' This
tells us straightaway that $\Gamma_z=8.07 \times 10^{-21}$ s, which is much longer than the
characteristic radiative damping time for the electron, $\Gamma_e =2e^2/3mc^3=6.26 \times
10^{-24}$ s (cf. eqn. 16.3 in Jackson [36]).

What can we claim to have accomplished with this approach? We have shown how a relation like
${\bf f}=m {\bf a}$ can be derived based on the electrodynamics of a ZPF-Planck oscillator
interaction. The electrodynamic parameter, $m_i$,  relating {\bf f} to {\bf a}
looks very much like an inertial mass. Assuming the ZPF-Planck oscillator interaction involves a
resonance, we can replicate the mass of the electron by choosing $\nu_r=\nu_C$ and $\Gamma_z
\nu_C=1$. Below we will discuss a physical argument for why $\nu_r$ must be the Compton
frequency, $\nu_C$. This argument involves a connection between this inertia-generating
resonance and the origin of the de Broglie wavelength of a moving electron. What remains to be
done is to establish some independent basis for determining $\Gamma_z$. It is tantalizing to
think that a physical understanding of the origin of $\Gamma_z$ for the electron might allow us
to predict additonal
$\Gamma_z$'s corresponding to excited resonances that might correspond to the muon and the
tauon, which appear to be simply heavy electrons.

Keeping in mind that only the electromagnetic interaction has been taken into account, the HRP
concept, if correct, substitutes for Mach's principle a very specific electromagnetic effect
acting between the ZPF and the charge inherent in matter. Inertia appears as an
acceleration-dependent electromagnetic (Lorentz) force. Newtonian mechanics would then be
derivable in principle from the ZPF via Maxwell's equations and in the more general case from
the other vacuum fields also. Note that this coupling of the electric and magnetic components of
the ZPF via the technique of Einstein and Hopf is very similar to that found in ordinary
electromagnetic radiation pressure. A similar observation, we conjecture, should hold for the
other vacuum fields. So we conclude that inertia appears as a radiation pressure exerted by the
fields in the vacuum opposing the acceleration of material elementary particles.

\bigskip\noindent
{\bf 4 \ \  The relativistic formulation of inertia from the ZPF Poynting Vector}
\bigskip

The oversimplification of an idealized oscillator
interacting with the ZPF as well as the mathematical complexity of the HRP analysis
are understandable sources of skepticism, as is the limitation to Newtonian
mechanics. A relativistic form of the equation of motion having standard covariant
properties has since been obtained [6] which is independent of any
particle model, since it relies solely on the standard Lorentz-transformation properties of
the electromagnetic fields.

Newton's third law states that if an agent applies a force to a point on an object, at
that point there arises an equal and opposite reaction force back upon the agent. In the case of
a fixed object  the equal and opposite reaction force can be traced to
interatomic forces in the neighborhood of the point of contact which act to resist
compression, and these in turn can be traced more deeply still to electromagnetic interactions
involving orbital electrons of adjacent atoms or molecules, etc.

Now a similar experience of an equal and opposite reaction force arises when a non-fixed object
is forced to accelerate. Why does acceleration create such a reaction force? We suggest that
this equal and opposite reaction force also has an underlying cause which is at least partially
electromagnetic, and specifically may be due to the scattering of ZPF radiation. Rueda \& Haisch
(RH) [6] demonstrated that from the point of view of the pushing agent there exists a net flux
(Poynting vector) of ZPF radiation transiting the accelerating object in a direction opposite to
the acceleration. The scattering opacity of the object to the transiting flux
would create a back reaction force that can be interpreted as inertia.

The RH approach is less complex and model-dependent than the HRP analysis in that it
assumes simply that electromagnetically-interacting elementary particles in any material object
interact with the ZPF in a way that produces ordinary electromagnetic scattering.
\footnote{$^e$}{It
is well known that treating the ZPF-particle interaction as dipole scattering is a successful
representation in that the dipole-scattered field exactly reproduces the original
unscattered field radiation pattern in unaccelerated reference frames.[16] It is thus
likely that dipole scattering is an appropriate way --- at least to first order --- to describe
several forms of ZPF-particle interaction.}
In the more general RH analysis one
simply needs to assume that there is some dimensionless efficiency factor,
$\eta(\omega)$, that describes whatever the process is. We suspect that $\eta(\omega)$
contains one or more resonances --- and in the following section discuss why these resonances
likely involve Compton frequencies of relevant particles forming a material object --- but again
this is not a necessary assumption.

The RH approach relies on making standard transformations of the
${\bf E}^{zp}$ and ${\bf B}^{zp}$ from a stationary to an accelerated coordinate
system. In a stationary or uniformly-moving frame the ${\bf E}^{zp}$ and
${\bf B}^{zp}$ constitute an isotropic radiation pattern. In an accelerated frame the
radiation pattern acquires asymmetries. There is thus a non-zero Poynting vector in
any accelerated frame carrying a non-zero net flux of electromagnetic momentum. The
scattering of this momentum flux generates a reaction force, ${\bf f}_r$. RH found that the
inertial mass is of the form

$$m_i={V_0 \over c^2} \int \eta(\nu) \rho_{zp}(\nu) \ d\nu , \eqno(8)
$$

\smallskip\noindent
where $\rho_{zp}$ is the well known spectral energy density of the ZPF  of eqn. (3).
\noindent The momentum of the object is of the form

$${\bf p}=m_i \gamma_{\tau} {\bf v}_{\tau} .
\eqno(9)$$

\noindent Not only does the ordinary form of Newton's second law, ${\bf f}=m{\bf a}$, emerge
from this analysis, but one can also obtain the relativistic form of the second law: [6]

$${\cal F}={d{\cal P} \over d\tau} = {d \over d\tau} (\gamma_{\tau} m_i c, \ {\bf p}
\ ) .
\eqno(10)$$

\indent The origin of inertia, in this picture, becomes remarkably intuitive. Any material object
resists acceleration because the acceleration produces a perceived flux of radiation
in the opposite direction that scatters within the object and thereby pushes against
the accelerating agent. Inertia in the present model appears as a kind of acceleration-dependent
electromagnetic vacuum-fields drag force acting upon electromagnetically-interacting elementary
particles.
The relativistic law for ``mass" transformation --- that is, the formula
describing how the {\it inertia} of a body has been calculated to change
according to an observer's relative motion --- is automatically satisfied
in this view, because the correct relativistic form of the reaction force
is derived, as shown in eqn (10).

\bigskip\noindent
{\bf 5 \ \ Inertial mass and the de Broglie relation for a moving particle: $\lambda = h/p$}
\bigskip
The four-momentum is defined as
%(Rindler, eqn. 27.4)
$$
{\bf P}= \left( {E \over c}, \ \ {\bf p} \right) = \left( \gamma m_0 c, \ \ {\bf p} \right)
=\left( \gamma m_0 c, \ \ \gamma m_0{\bf v} \right), \eqno(11)$$
where $|{\bf P}|=m_0 c$ and $E=\gamma m_0 c^2$. The Einstein-de Broglie relation defines the
Compton frequency
$h \nu_C = m_o c^2$ for an object of rest mass $m_0$, and if we make the de Broglie
assumption that the momentum-wave number relation for light also characterizes matter
then ${\bf p}=\hbar {\bf k}_B$ where ${\bf
k}_B=2\pi(\lambda^{-1}_{B,1},\lambda^{-1}_{B,2},\lambda^{-1}_{B,3})$. We thus write
$$
{{\bf P} \over \hbar} = \left( {2\pi \gamma \nu_C \over c}, {\bf k}_B \right)
= 2 \pi \left( {\gamma \over \lambda_C}, {1 \over \lambda_{B,1}}, {1 \over \lambda_{B,2}}, {1
\over
\lambda_{B,3}} \right)
\eqno(12)
$$
and from this obtain the relationship
$$
\lambda_B={c \over \gamma v} \lambda_C \eqno(13)
$$
between the Compton wavelength, $\lambda_c$, and the de Broglie wavelength, $\lambda_B$. For a
stationary object $\lambda_B$ is infinite, and the de Broglie wavelength decreases in inverse
proportion to the momentum.

Eqns. (6), (7) and (8) are very suggestive that ZPF-elementary particle interaction involves a
resonance at the Compton frequency. 
De Broglie proposed that an elementary particle is associated with a localized
wave whose frequency is the Compton frequency.
As summarized by Hunter [37]: ``\dots what we regard as the (inertial) mass of the particle
is, according to de Broglie's proposal, simply the vibrational energy (divided by $c^2$)
of a localized oscillating field (most likely the electromagnetic field). From this
standpoint inertial mass is not an elementary property of a particle, but rather a
property derived from the localized oscillation of the (electromagnetic) field. De Broglie
described this equivalence between mass and the energy of oscillational motion\dots as
{\it `une grande loi de la Nature'} (a great law of nature).'' 

This perspective is
consistent with the proposition that inertial mass, $m_i$, may be a coupling parameter between
electromagnetically interacting particles and the ZPF. Although De Broglie assumed
that his wave at the Compton frequency originates in the particle itself (due to some
intrinsic oscillation or circulation of charge perhaps) there is an alternative interpretation
discussed in some detail by de la Pe\~na and Cetto that a particle  ``is tuned to a wave
originating in the high-frequency modes of the zero-point background field.''[38] The de Broglie
oscillation would thus be due to a resonant interaction with the ZPF, presumably the same
resonance that is responsible for creating a contribution to inertial mass as in eqns. (7)
and (8). In other words, the ZPF would be driving this
$\nu_C$ oscillation.

We therefore suggest that an elementary charge driven to oscillate at the Compton
frequency, $\nu_C$,  by the ZPF may be the physical basis of the $\eta(\nu)$
scattering parameter in eqn. (8).  For the case of the electron, this would imply that
$\eta(\nu)$ is a sharply-peaked resonance at the frequency, expressed in terms of
energy, $h\nu_C=512$ keV. The inertial mass of the electron would physically be the reaction
force due to resonance scattering of the ZPF at that frequency.

{\it This leads to a surprising corollary.} It has been shown that as viewed from a
laboratory frame, a standing wave at the Compton frequency in the electron frame transforms
into a traveling wave having the de Broglie wavelength
for a moving electron.[4][37][38][39] The wave nature of the moving electron (as measured in the
Davisson-Germer experiment, for example) would be basically due to Doppler shifts associated with
its Einstein-de Broglie resonance at the Compton frequency.
A simplified heuristic model shows this, and a detailed treatment showing the same result
may be found in de la Pe\~na and Cetto [4]. Represent a ZPF-like driving force field as two
waves having the Compton frequency $\omega_C=2\pi \nu_C$ travelling in equal and opposite
directions,
$\pm
\hat{x}$. The amplitude of the combined wave acting upon an electron fixed at a given coordinate
$x$ will be
$$
\phi=\phi_+ + \phi_{-}=2 \cos \omega_C \cos k_C . \eqno(14)
$$

But now assume an electron is moving with velocity $v$ in the $+x$-direction. The wave
responsible for driving the resonant oscillation impinging on the electron from the front
will be the wave seen in the laboratory frame to have frequency $\omega_-=\gamma
\omega_C (1 - v/c)$, i.e. it is the wave below the Compton frequency in the laboratory
that for the electron is Doppler shifted up to the
$\omega_C$ resonance. Similarly the ZPF-wave responsible for driving the electron resonant
oscillation impinging on the electron from the rear will have a laboratory frequency
$\omega_+=\gamma \omega_C (1 + v/c)$ which is Doppler shifted down to $\omega_C$ for the
electron. The same transformations apply to the wave numbers,
$k_+$ and $k_-$. The Lorentz invariance of the ZPF spectrum ensures that regardless of the
electron's (unaccelerated) motion the up- and down-shifting of the laboratory-frame
ZPF will always yield a standing wave in the electron's frame.

It can be shown [4][39] that the superposition of these two waves is
$$
\phi'=\phi'_++\phi'_{-}=2 \cos(\gamma \omega_C t - k_B x) \cos(\omega_B t - \gamma k_C x).
\eqno(15)
$$
Observe that for fixed $x$, the rapidly
oscillating ``carrier'' of frequency $\gamma \omega_C$ is modulated by the slowly varying
envelope function in frequency $\omega_B$. And {\it vice versa} observe that at a given $t$ the
``carrier'' in space appears to have a relatively large wave number $\gamma k_C$ which is
modulated by the envelope of much smaller wave number $k_B$. Hence
both timewise at a fixed point in space and spacewise at a given time, there appears a
carrier that is modulated by a much broader wave of dimension corresponding to the de
Broglie time $t_B=2\pi/\omega_B$, or equivalently, the de Broglie wavelength
$\lambda_B=2\pi/k_B$.

This result may be generalized to include ZPF radiation from all other directions, as may
be found in the monograph of de la Pe\~na and Cetto [4]. They conclude by stating:
``The foregoing discussion assigns a physical meaning to de Broglie's wave: it is the {\it
modulation} of the wave formed by the Lorentz-transformed, Doppler-shifted superposition
of the whole set of random stationary electromagnetic waves of frequency
$\omega_C$ with which the electron interacts selectively.''

Another way of looking at the spatial modulation is in terms of the
wave function: the  spatial modulation of eqn. (15) is exactly the $e^{i p x / \hbar}$ wave
function of a freely moving particle satisfying the Schr\"odinger equation.
The same argument has been made by Hunter [37].
In such a view the quantum wave function
of a moving free particle becomes a ``beat frequency'' produced by the relative motion of
the observer with respect to the particle and its oscillating charge.

It thus appears that a simple model of a particle as a ZPF-driven oscillating charge
with a resonance at its Compton frequency may simultaneously offer insight into the nature
of inertial mass, i.e. into rest inertial mass and its relativistic extension, the
Einstein-de Broglie formula and into its associated wave function involving the de Broglie
wavelength of a moving particle. If the de Broglie oscillation is indeed driven by the ZPF,
then it is a form of Schr\"odinger's {\it zitterbewegung}. Moreover there is a substantial
literature attempting to associate spin with {\it zitterbewegung} tracing back to the work
of Schr\"odinger [12]; see for example Huang [13] and Barut and Zanghi [40].

\bigskip\noindent
{\bf 6 \ \ Comments on Gravitation}
\bigskip
If inertial mass, $m_i$, originates in quantum vacuum-charge type interactions, then, by the
principle of equivalence so must gravitational mass, $m_g$. In this view, and within the
restricted context of the electromagnetic approach, gravitation would be a force originating in
ZPF-charge interactions analogous to the ZPF-inertia concept. Sakharov [9], inspired by previous
work of Zeldovich [41], was the first to conjecture in a more general sense this interpretation
of gravity. If true, gravitation would be unified with the other forces: it would be a
manifestation of the other fields.

General relativity (GR) attributes gravitation to spacetime curvature. Modern attempts to
reconcile quantum physics with GR take a different approach, treating gravity as an exchange of
gravitons in flat spacetime (analagous to the treatment of electromagnetism as exchange of
virtual photons). A non-geometric (i.e. flat spacetime) approach to gravity is legitimate in
quantum gravity. Similarly another non-geometric approach would be to assume that the dielectric
properties of space itself may change in the presence of matter: this can be called the
polarizable vacuum (PV) approach to gravity. Propagation of light in the presence of matter
would deviate from straight lines due to variable refraction of space itself, and other GR
effects such as the slowing down of light (the coordinate velocity as judged by a distant
observer) in a gravitational potential would also occur. But of course it is the propagation of
light from which we infer that spacetime is curved in the first place. This raises the
interesting possibility that GR may be successful and yet not because spacetime is really
curved: rather because the point-to-point changes in the dielectric (refractive) properties of
space in the presence of matter create the illusion of geometrical curvature. A PV type of model
does not directly relate gravitation to the ZPF (or to the more general quantum vacuum) but it
does appear to provide a theoretical framework conducive to developing the conjecture of Sakharov
that it is changes in the ZPF that create gravitational forces.

There were some early pioneering attempts, inspired by Sakharov's conjecture, to link gravity
to the vacuum from a quantum field theoretical viewpoint  (by Amati, Adler and others, see
discussion and references in Misner, Thorne and Wheeler [42]) as well as within SED (see
Surdin [43]). The first step in developing Sakharov's conjecture in any detail within the
classical context of nonrelativistic SED was the work of Puthoff [11].  In this approach
gravity is treated as a residuum force in the manner of the van der Waals forces. Expressed in
the most rudimentary way this can be viewed as follows. The electric component of the ZPF causes
a given charged particle to oscillate. Such oscillations give rise to secondary electromagnetic
fields. An adjacent charged particle will thus experience both the ZPF driving forces causing
it to oscillate, and in addition forces due to the secondary fields produced by the ZPF-driven
oscillations of the first particle. Similarly, the ZPF-driven oscillations of the second
particle will cause their own secondary fields acting back upon the first particle. The net
effect is an attractive force between the particles. The sign of the charge does not matter:
it only affects the phasing of the interactions. Unlike the Coulomb force which, classically
viewed, acts directly between charged particles, this interaction is mediated by extremely
minute propagating secondary fields created by the ZPF-driven oscillations, and so is
enormously weaker than the Coulomb force.  Gravitation, in this view, appears to be a
long-range interaction akin to the van der Waals force.

The Puthoff analysis consists of two separate parts. In the first, the energy
of Schr\"odinger's {\it zitterbewegung} motion is equated to gravitational mass, $m_g$ (after
dividing by $c^2$). This leads to a relationship between $m_g$ and electrodynamic parameters
that is identical to the HRP inertial mass, $m_i$, apart from a factor of two. This factor of
two is discussed in the appendix of HRP, in which it is concluded that the Puthoff $m_g$
should be reduced by a factor of two, yielding $m_i=m_g$ precisely.

The second part of Puthoff's analysis is more controversial. He  quantitatively examines the
van der Waals force-like interactions between two driven oscillating dipoles and derives an
inverse square force of attraction. This part of the analysis has been challenged by Carlip
to which Puthoff has responded [44], but, since problems remain [45], this aspect of the
ZPF-gravitation concept requires further theoretical development, in particular the
implementation of a fully relativistic model.

One might think that the ZPF-inertia and the ZPF-gravitation concepts must
stand or fall together, given the principle of equivalence. Yet this may
not be the case. Following the notation of Jammer [1] one may identify two
aspects of gravitational mass: $m_a$, the active gravitational mass, is
the source of gravitational field, and $m_p$, the passive gravitational
mass, describes a body's response to an imposed gravitational field. In a
geometrical theory of gravity $m_i = m_p$, the principle of equivalence,
is automatically satisfied, because a gravitational ``force'' {\it is} an
inertia reaction force as seen in the locally Minkowskian frame. This
remains true whether $m_i$ is intrinsic or a product of extrinsic
interactions. The identity $m_i = m_a$ is not as immediately obvious, but
since the source of gravitational curvature is energy (or rather, the
energy-momentum tensor), $m_i = m_a$ will follow as long as the
ZPF-inertia theory respects the relativistic relationship between inertia
and energy, as discussed in earlier sections.

Although ZPF-inertia does not require ZPF-gravity as a support, it is the
case that a ZPF-driven theory of gravity such as the one attempted by
Puthoff would legitimately refute the objection that the ZPF cannot be a
real electromagnetic field since the energy density of this field would be
enormous and thereby act as a cosmological constant, $\Lambda$, of
enormous proportions that would curve the Universe into something
microscopic in size. This cannot happen in the Sakharov-Puthoff view. This situation is clearly
ruled out by the fact that, in this view, the ZPF cannot act upon itself to gravitate.
Gravitation is not caused by the mere presence of the ZPF, rather by secondary motions of
charged particles driven by the ZPF. {\it In this view it is impossible for the ZPF to give
rise to a cosmological constant.} (The possibility of non-gravitating vacuum energy has
recently been investigated in quantum cosmology in the framework of the modified
Born-Oppenheimer approximation by Datta [46].)

The other side of this argument is of course that as electromagnetic radiation is not made of
polarizable entities one might naively no longer expect deviation of light rays by massive
bodies. We speculate however that such deviation will be part of a fully relativistic theory
that besides the ZPF properly takes into account the polarization of the Dirac vacuum when
light rays pass through the particle-antiparticle Dirac sea. It should act, in effect, as a
medium with an index of refraction modified in the vicinity of massive objects. This is very
much in line with the original Sakharov [9] concept. Indeed, within a more general
field-theoretical framework one would expect that the role of the ZPF in the inertia and
gravitation developments mentioned above will
be played by a more general quantum vacuum field, as was already suggested in the HRP appendix.

\bigskip\noindent
{\bf 7 \ \ Concluding comments on the Higgs Field as originator of mass}
\bigskip

In the Standard Model of particle physics it is postulated that there exists a scalar field
pervasive throughout the Universe and whose main function is to assign mass to the elementary
particles. This is the so-called Higgs field or Higgs boson and it originated from a proposal by
the British physicist Peter Higgs who introduced that kind of field as an idea for assigning
masses in the Landau-Ginzburg theory of superconductivity. Recent predictions of the mass that
the Higgs boson itself may have indicate a rather large mass (more than 60 GeV) and this may be
one of the reasons why, up to the present, the Higgs boson has not been observed. There are
alternative theories that give mass to elementary particles without the need to postulate a Higgs
field, as, e.g., dynamical symmetry breaking where the Higgs boson is not elementary but
composite. But the fact that the Higgs boson has not been detected is by no means an indication
that it does not exist. Recall the 26 years which passed between the proposal by Pauli in 1930
of the existence of the neutrino and its first detection when the Reines
experiment was performed.

It should be clearly stated that the existence (or non-existence) of the hypothetical Higgs boson
does not affect our proposal for the origin of inertia. In the Standard Model attempt to
obtain, in John Wheeler's quote, ``mass without mass,'' the issue of inertia itself does not
appear. As Wilczek [47] states concerning protons and neutrons: ``Most of the mass of ordinary
matter, for sure, is the pure energy of moving quarks and gluons. The remainder, a
quantitatively small but qualitatively crucial remainder --- it includes the mass of electrons
--- is all ascribed to the confounding influence of a pervasive medium, the Higgs field
condensate.'' An explanation of proton and neutron masses in terms of the energies of quark
motions and gluon fields falls short of offering any insight on inertia itself. One is no closer
to an understanding of how this energy somehow acquires the property of resistance to
acceleration known as inertia. Put another way, a quantitative equivalence between energy and
mass does not address the origin of inertial reaction forces.

Many physicists apparently believe that our conjecture of inertia originating in the vacuum
fields is at odds with the Higgs hypothesis for the origin of mass. This happens because of the
pervasive, one might even say invisible, assumption that inertia can only be intrinsic to mass
and thus if the Higgs mechanism creates mass one automatically has an explanation for
inertia. If inertia is intrinsic to mass as postulated by Newton, then inertia could indeed be
considered to be a direct result of the Higgs field because presumably the Higgs field is the
entity that generates the corresponding mass and inertia simply comes along with mass
automatically. However if one accepts that there is indeed an extrinsic origin for the inertia
reaction force, be it the gravity field of the surrounding matter of the Universe (Mach's
Principle) or be it the electromagnetic quantum vacuum (or more generally the quantum vacua)
that we propose, then the question of how mass originates --- possibly by a Higgs mechanism ---
is a separate issue from the property of inertia. This is a point that is often not properly
understood. The modern Standard Model explanation of mass is satisfied if it can balance the
calculated energies with the measured masses (as in the proton) but obviously this does not
explain the origin of the inertia reaction force. Returning to our {\it epistemology of
observables}, it is the inertia reaction force associated with acceleration that is measureable
and fundamental, not mass itself. We are proposing a specific mechanism for generation of the
inertia reaction force resulting from distortions of the quantum vacua as perceived by
accelerating elementary particles.

We do not  enter into the problems associated with attempts to explain inertia via Mach's
Principle, since we have discussed this at length in a recent paper [48]: a detailed discussion
on intrinsic vs. extrinsic inertia and on the inability of the geometrodynamics of general
relativity to generate inertia reaction forces may be found therein. It had already been shown
by Rindler [49] and others that Mach's Principle is inconsistent with general relativity, and
Dobyns et al. [48] further elaborate on a crucial point in general relativity that is not much
appreciated: Geometrodynamics merely defines the geodesic that a freely moving object will
follow. But if an object is constrained to follow some different path, geometrodynamics has no
mechanism for creating a reaction force. Geometrodynamics has nothing more to say about inertia
than does classical Newtonian physics. Geometrodynamics leaves it to whatever processes generate
inertia to generate such a force upon deviation from a geodesic path, but this becomes an obvious
tautology if an explanation of inertia is sought in geometrodynamics.

Concerning neutrino mass,
if, unlike the neutron which consists of three quarks whose charges cancel, the
neutrino is truly a neutral particle, it could have no electromagnetically originating mass. It
was announced in 1997 that the Super-Kamiokande neutrino observatory had, at last, succeeded in
measuring a mass for the neutrino. Of course these measurements did
not directly measure the property of inertial mass; this is an impossibility at present. What
was measured was the ratio of electron neutrinos to muon neutrinos due to cosmic rays. In the
current Standard Model of particle physics, this ratio implies an oscillation between the two
types of neutrinos which in turn implies a theoretical mass. This is a ``mass'' based on a
specific interpretation from the Standard Model not a direct measurement of inertial mass (and
the quantum vacuum-inertia concept of mass proposes specifically that mass is a quite different
thing than the concept of mass in the Standard Model).  However there is a more likely
resolution. There are two other vacuum fields: those associated with the weak and strong
interactions. The neutrino is governed by the weak
interaction, and it is possible that a similar kind of ZPF-particle interaction creates inertial
mass for the neutrino but now involving the ZPF of the weak interaction. At present this is pure
conjecture. No theoretical work has been done on this problem. In either case, it is prudent to
be open to the possibility that certain areas of standard theory may benefit from a fundamental
reinterpretation of mass which would resolve these apparent conflicts.

Inertia is frequently taken as the defining feature of mass in the
development of classical and relativistic mechanics. This has the virtue
of parsimony, but a deeper understanding of the profound connections
between inertia and energy, and inertia and gravity, may be achievable if
a consistent theory for a dynamical origin of inertia can be found. The
question of why the mass associated with either matter or energy should
display a resistance to acceleration is a valid one that needs to be
addressed even if the Higgs boson is experimentally found and confirmed as
the origin of mass. 

\bigskip
{\it Acknowledgement ---} We acknowledge NASA contract NASW-5050 for support of this research.

\bigskip
{

\bigskip

\parskip=0pt plus 2pt minus 1pt\leftskip=0.25in\parindent=-.25in 

\centerline{REFERENCES}

[1] M. Jammer, Concepts of Mass in Contemporary Physics and Philosopy, Princeton Univ. Press
(2000)

[2] A. Einstein, Ann. der Physik {\bf 15} (1905) xxxx; but see Dirac's comments on the need for a redefined ether
in P.A.M. Dirac, Nature, {\bf 168} (1951) 906

[3] P. Frank, in Albert Einstein: Philosopher-Scientist, (P. A. Schilp, ed.), Vol. I, 269. (1959)

[4] L. de la Pe\~na, and A. M. Cetto, The Quantum Vacuum: An Introduction to Stochastic
Electrodynamics, Kluwer Acad. Publ. (1996)

[5] B. Haisch, A. Rueda, and H. E. Puthoff, Phys. Rev. A {\bf 49} (1994) 678 (HRP)

[6] A. Rueda and B. Haisch, Found. Phys. {\bf 28}
(1998) 1057; A. Rueda and B. Haisch, Phys. Lett. A {\bf 240} (1998) 115; 

[7] J.-P. Vigier, Found. Phys. {\bf 25}, 1461 (1995).

[8] M.-T. Jaekel and S. Raynaud, 1995, Quantum and Semiclassical Optics, {\bf 7} (1905), 499;
M.-T. Jaekel and S. Raynaud, Repts. Prog. Phys., {\bf 60} (1997) 863; also reference therein to
relevant work of other authors.

[9] A. D. Sakharov, Dokl. Akad. Nauk. SSR (Sov. Phys. Dokl.) {\bf 12} (1968) 1040

[10] D. Hestenes, Found. Phys. {\bf 20}, 1213 (1990) and D. Hestenes, ``Zitterbewegung in
Radiative Processes'' in The Electron -- New Theory and Experiment. D. Hestenes and A.
Weingartshoer, eds. (Kluwer, Dordrecht, 1991).

[11] H. E. Puthoff,  Phys. Rev. A, {\bf 39} (1989) 2333

[12] E. Schr\"odinger, Sitz. Ber. Preuss. Akad. Wiss. Phys.-Math. Kl, {\bf 24}, (1930) 4318

[13] K. Huang, Am. J. Phys. {\bf 20} (1952) 479

[14] D. C. Cole, A. Rueda, and K. Danley (2000), in preparation; K. Danley, (1994),
unpublished thesis, Calif. State Univ. Long Beach.

[15] R. Loudon, The Quantum Theory of Light (2nd ed.) Clarendon Press, Oxford (1983)

[16] T. H. Boyer, Phys. Rev. D, {\bf 11} (1975) 790

[17] T. W. Marshall, Proc. Roy. Soc. London, Ser. A, {\bf 276} (1963) 475; Proc. Cambridge Phil.
Soc. {\bf 61} (1965) 537

[18] M. Planck, Ann. Physik {\bf 4} (1901) 553; M. Planck, Verhandl. Deutsch. Phys. Ges.
{\bf 13} (1911) 138

[19] T. Kuhn, T. Black Body Theory and the Quantum Discontinuity: 1894--1912, (Oxford Univ.
Press, Oxford) (1978)

[20] A. Einstein and O. Stern, Ann. Physik {\bf 40} (1913) 551

[21] W. Nernst, Verhandlungen der Deutschen Physikalischen Gesellschaft {\bf 4} (1916) 83

[22] T.H. Boyer,Phys. Rev. D {\bf 29} (1984)  1096; also Phys Rev. D, {\bf 29} (1984) 1089

[23] N. Bohr, N. Phil. Mag.  {\bf 26}, No. 1, (1913) 476, 857.

[24] H. E. Puthoff,  Phys. Rev. D, {\bf 35} (1987) 3266

[25] P. Claverie, L. Peqquera, F. Soto, Phys. Lett. A {\bf 80}, 113; and references therein.

[26] D. C. Cole, Found. Phys. {\bf 20}, (1990) 225.

[27] A. Rueda, Found. Phys. Lett. {\bf 6}, No. 1 (1993), 75; {\bf 6} No. 2 (1993) 139

[28] B. Haisch, A. Rueda, and H. E. Puthoff, Spec. in Sci. and Tech. {\bf 20} (1987) 99

[29] S. Hawking, Nature, {\bf 248} (1974) 30

[30] P. C. W. Davies, J. Phys. A, {\bf 8} (1975) 609

[31] W. G. Unruh, Phys. Rev. D, {\bf 14} (1976) 870

[32] P. C. W. Davies, T. Dray, and C. A. Manogue, Phys.Rev. D {\bf 53} (1996) 4382

[33] T. H. Boyer, Phys. Rev. D {\bf 21} (1980) 2137

[34] T. H. Boyer, Phys. Rev. D {\bf 29} (1984) 1089

[35] A. Einstein, and L. Hopf, Ann. Phys., {\bf 33} (1910) 1096; {\bf 33} (1910) 1105

[36] J. D. Jackson, Classical Electrodynamics (3rd ed.) (1975)

[37] G. Hunter, in The Present Status of the Quantum Theory of Light, S. Jeffers et al. (eds.),
(Kluwer Acad. Publ.), chap. 12 (1996)

[38] A. F. Kracklauer, Physics Essays {\bf 5} (1992) 226; for a formal derivation and further
illuminating discussion see Chap. 12 of [4]

[39] B. Haisch and A. Rueda, Phys. Lett. A, {\bf 268}, (2000) 224.

[40] A. O. Barut and N. Zanghi, Phys. Rev. Lett. {\bf 52} (1984) 209

[41] Ya. B. Zeldovich, ZhETF Pis'ma {\bf 6}, (1967) 922, 1233; JETP Lett. {\bf 6} (1967) 345

[42] C. W. Misner, K. S. Thorne, and J. A. Wheeler,(1973), Gravitation, (W.H. Freeman, San
Francisco).

[43] M. Surdin, Found. Phys. {\bf 8} (1978) 341

[44] S. Carlip, Phys. Rev. A, {\bf 47} (1993) 3452; H. E. Puthoff, Phys. Rev. A {\bf 47} (1993)
3454.

[45] D. C. Cole, A. Rueda and K. Danley, (2000), in preparation.

[46] D. P. Datta, Class. Quantum Grav. {\bf 12} (1995) 2499

[47] Wilczek, Physics Today, Nov. 1999, p.. 11 and Jan. 2000, p. 13

[48] Y. Dobyns, A. Rueda and B. Haisch, Found. Phys. (2000) {\bf 30}, No. 1, 59

[49] W. Rindler, Phys. Lett. A {\bf 187} (1994) 236; W. Rindler, Phys. Lett. A {\bf 233} (1997)
25

}

\bye